\documentclass[aps,prl,twocolumn,groupedaddress]{revtex4-1}

\usepackage{outlines}
\usepackage{dsfont}
\usepackage{enumitem}
\usepackage{braket}
\usepackage{amsmath}
\usepackage{mathtools}
\usepackage{amsfonts,amssymb,amsbsy}
\usepackage{graphicx}
\usepackage{MnSymbol}
\usepackage{xcolor}
\usepackage{hyperref}

\hypersetup{
    colorlinks=true,
    linkcolor=blue,
    filecolor=magenta,      
    urlcolor=red,
}

\begin{document}

%\preprint{}

\title{Eliminating Leakage Errors in Hyperfine Qubits}


\author{D. Hayes}
\email{david.hayes@honeywell.com}
\author{D. Stack}
\author{B. Bjork}
\author{A. C. Potter}
\author{C. H. Baldwin}
\author{R. P. Stutz}
\affiliation{Honeywell Quantum Solutions}

\date{\today}

\begin{abstract}
Population leakage outside the qubit subspace presents a particularly harmful source of error that cannot be handled by standard error correction methods. Using a trapped $^{171}$Yb$^+$ ion, we demonstrate an optical pumping scheme to suppress leakage errors in atomic hyperfine qubits. The selection rules and narrow linewidth of a quadrupole transition are used to selectively pump population out of leakage states and back into the qubit subspace. Each pumping cycle reduces the leakage population by a factor of $\sim3$, allowing for an exponential suppression in the number of cycles. We use interleaved randomized benchmarking on the qubit subspace to show that this pumping procedure has negligible side-effects on the qubit subspace, bounding the induced qubit memory error by $\leq2.0(8)\times10^{-5}$ per cycle, and qubit population decay to $\leq1.4(3)\times10^{-7}$ per cycle. These results clear a major obstacle for implementations of quantum error correction and error mitigation protocols.
\end{abstract}

%\keywords{}

\maketitle

Qubits are the starting point in most quantum computing architectures. Idealized two-level systems are ubiquitous in theoretical proposals and analysis, but are commonly only approximately realized in experiments by restricting the relevant dynamics to just two levels. This paradigm has been the cornerstone of demonstrations of high fidelity qubit initialization, quantum memories, single-qubit and two-qubit gates, state-detection\cite{Langer05,Olmschenk07,Benhelm08,Harty14,Ballance16,Gaebler16,Wang17}, and algorithms of increasing complexity\cite{Gulde03,Brickman05,Debnath16,Linke17}. Deviations from these idealized models are known as leakage errors, which are quantum processes that drive population from a qubit subspace to other levels supported by the physical medium. While the experimental progress in quantum information processing is evident, the ultimate noise sensitivity of these machines is notoriously difficult to predict, and leakage errors are perhaps amongst the most worrisome.

Powerful algorithms require large circuit depths, necessitating the use of error correction and mitigation techniques. One technique of particular importance is quantum error correction, which allows for the efficient suppression of errors to an arbitrary level, thereby allowing for the possibility of universal quantum computation\cite{Nielsen00}. However, proofs of fault tolerance via quantum error correcting codes commonly assume the errors and subsequent corrections act within the qubit space and studies have shown that leakage can have a devestating effect on these codes\cite{Ghosh13}. Other error mitigation techniques circumvent active feedback and are more akin to open-loop control; including random sampling of noisy circuits\cite{Temme17}, quantum subspace expansion\cite{McClean17}, and stabilizer based error mitigation\cite{Bonet18}. These methods are attractive for near-term use since they do not require large qubit overheads, but they too assume noise models that act on qubits and may be ineffective on systems with significant leakage errors. 

Researchers have recognized this issue and constructed various leakage reducing units\cite{Suchara14}, but they typically require extra qubit resources and circuitry, consequently lowering the physical error rate threshold. This addition to the already formidable overhead of quantum error correction has even led some researchers to consider abandoning hyperfine clock-qubits and their magnetic field insensitivity in favor of a leakage-free qubit\cite{Brown18}. 

In this work, we propose a scheme to suppress leakage errors at the physical level in hyperfine clock-qubits. Namely, we propose and demonstrate an optical pumping scheme that drives leaked population incoherently back into the qubit subspace, thereby converting leakage errors into conventional qubit errors, which can be handled by error correction and mitigation techniques. Each leakage repump pulse succeeds probabilistically, resulting in an exponential reduction of leakage population with pulse number. Utilizing atomic selection rules and pulse shaping techniques, we ensure that this pumping has a negligible impact on the qubit subspace.These results bring trapped-ion platforms in line with the assumptions of error-correction and mitigation techniques.

We analyze and demonstrate our leakage suppression scheme in $^{171}$Yb$^+$. This system has a simple hyperfine structure and a low-field clock-qubit in the $^2S_{1/2}$ manifold, $\{\ket{F=0,m_f=0},\ket{F=1,m_f=0}\}\equiv\{\ket{0},\ket{1}\}$, and can be controlled using stimulated Raman transitions through the excited $^2P_{1/2}$ and $^2P_{3/2}$ states. Spontaneous scattering during these gates is a fundamental error that is partially leakage inducing\cite{Ozeri07}. In the case of $^{171}$Yb$^+$, these scattering events drive population into two leakage states, $\ket{L_{\pm}}\equiv{}^2S_{1/2}\ket{F=1,m_f=\pm1}$. In principle, population in these states can be optically pumped back into the qubit space through the dipole transition $\ket{L_{\pm}}\leftrightarrow{}^2P_{1/2}\ket{F=0,m_f=0}$ using $\hat{\sigma}_{\mp}$ light. However, this scheme's usefulness is fundamentally limited by off-resonant transitions $\ket{1}\leftrightarrow{}^2P_{1/2}\ket{F=1,m_f=\pm1}$. Moreover, even small $\hat{\pi}$-polarization impurities at the $10^{-4}$ level would drive the $\ket{1}\leftrightarrow{}^2P_{1/2}\ket{F=0,m_f=0}$ transition and cause a $\mathcal{O}(10^{-2})$ qubit error per scattering event, meaning that a pumping procedure requiring several scattering events would induce a large error. Instead, we propose a much more robust scheme that takes advantage of the narrow linewidth and selection rules of the $^2S_{1/2}\leftrightarrow{}^2D_{3/2}$ quadrupole transition.

\begin{figure}
  \centering
    \includegraphics[width=0.4\textwidth]{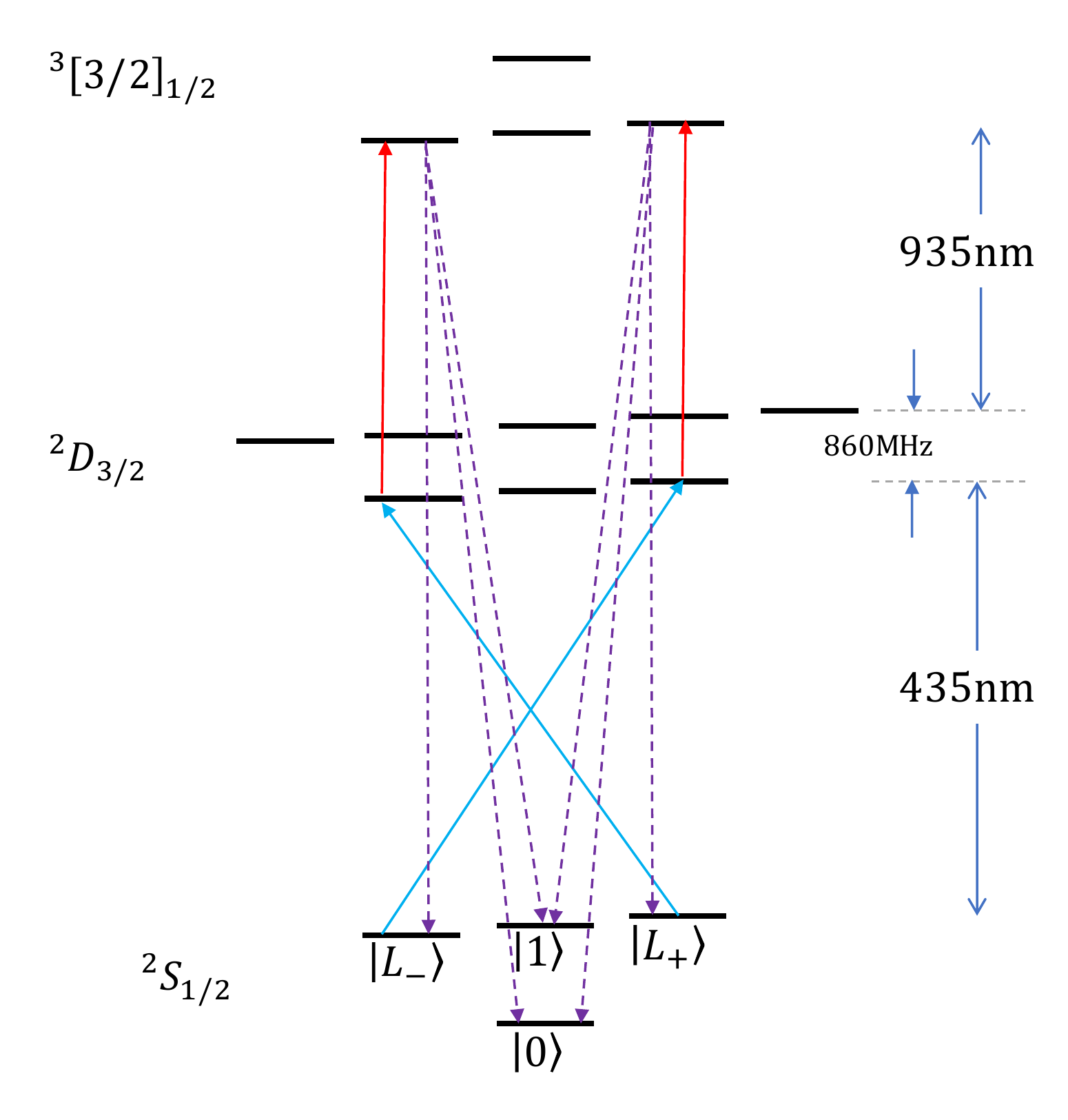}
\caption{A level diagram illustrating the leakage repump scheme. The solid blue lines show the 435nm quadrupole transition driving leaked population from $\ket{L_{\pm}}$ to $^2D_{3/2}$. The second step, illustrated by solid red lines, is a dipole transition at 935nm that drives population to $^3[3/2]_{1/2}$. As illustrated by the dashed lines, $^3[3/2]_{1/2}$ states decay to $^2S_{1/2}$ with high probability, thus completing the repump cycle.}
\label{fig:Levels}
\end{figure}

The quadrupole transition Rabi frequency between angular momentum states with z-components $m_{f_1}, m_{f_2}$ is proportional to a geometric factor $g^{(\Delta m_f)}$~\cite{BenhelmThesis}, given by,
\begin{align}
g^{(0)}(\theta,\phi)&=\frac{1}{2}|\cos \theta \sin 2\phi| \\
g^{(1)}(\theta,\phi)&=\frac{1}{\sqrt{6}}|\cos \theta \cos 2\phi+i \sin \theta \cos \phi| \\
g^{(2)}(\theta,\phi)&=\frac{1}{\sqrt{6}}|\frac{1}{2}\cos \theta \sin 2\phi+i \sin \theta \sin \phi|,
\end{align}
where $\theta$ and $\phi$ are the respective polarization and k-vector angles relative to the quantization axis defined by the local magnetic field. When the polarization and k-vector are both orthogonal to the quantization axis, $g^{(0)}(\pi/2,\pi/2)=g^{(1)}(\pi/2,\pi/2)=0$ and $g^{(2)}(\pi/2,\pi/2)=1/\sqrt{6}$, meaning that only $\Delta m_f=\pm2$ transitions will occur as illustrated in Fig.~\ref{fig:Levels}. When the laser is tuned into resonance  with the $^2S_{1/2}\ket{F=1}\leftrightarrow{}^2D_{3/2}\ket{F=1}$ transitions, the selection rules and $^2D_{3/2}$ hyperfine splitting ensure that only the leakage states will be transferred to the $^2D_{3/2}$ manifold, leaving the qubit states unperturbed. After the leakage population has been transferred to the $^2D_{3/2}$ state, we apply $\hat{\pi}$-polarized 935nm light that is resonant with the $^2D_{3/2}\ket{F=1}\leftrightarrow{}^3[3/2]_{1/2}\ket{F=1}$ transition\cite{Olmschenk07}. The $^3[3/2]_{1/2}$ state has a 38ns lifetime and returns the ion to $^2S_{1/2}$ with a $98.2\%$ branching ratio. Assuming perfect quadrupole transfer pulses and ignoring the $1.8\%$ decay back to $^2D_{3/2}$, the leaked population will be reduced as $P_0\rightarrow P_0/3$ per cycle. We note that the $1.8\%$ decay to $^2D_{3/2}\ket{F=2}$ can be mitigated by adding a second frequency or by power broadening. After $n$ cycles, the leaked population would be reduced as $P_0\rightarrow P_0/3^n$ and we note that polarization errors in the $935$nm light would, at worst, lead to a $P_0\rightarrow P_0(2/3)^n$ reduction.

\begin{figure}
  \centering
    \includegraphics[width=0.45\textwidth]{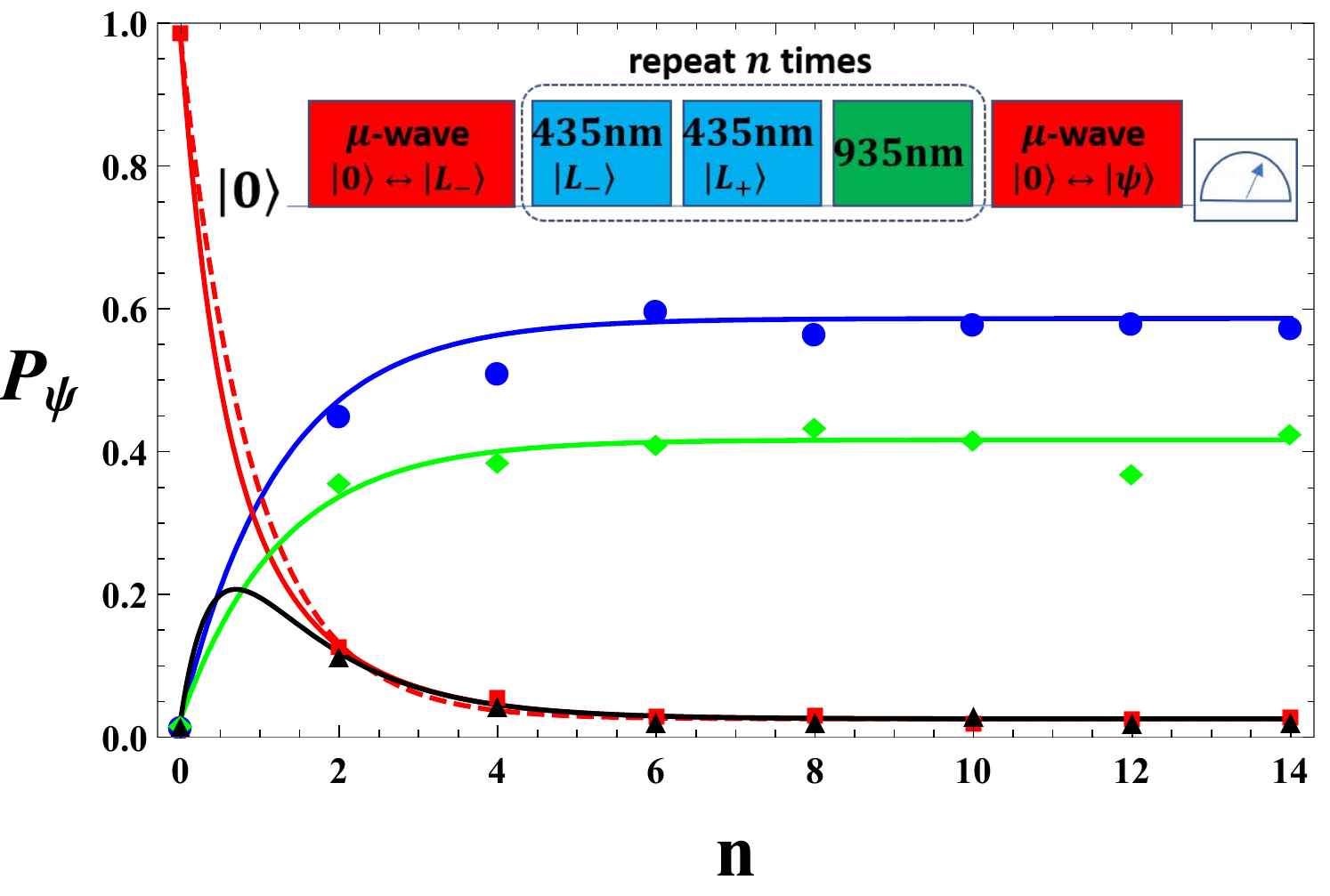}
\caption{Pumping a leaked state $\ket{L_-}$ into the qubit subspace. Qubit populations, $P_0$ and $P_1$, are shown in \textcolor{blue}{blue} circles and \textcolor{green}{green} diamonds respectively and leakage populations, $P_{L_-}$ and $P_{L_+}$, are shown in \textcolor{red}{red} squares and \textcolor{black}{black} triangles respectively. Each point is the average of 1000 shots with a  standard error on the $1\%$ level. The solid lines show fitted curves described in the main text and the dashed line shows the ideal decay of $1/3^n$. The pulse sequence is also shown, and we've omitted the final 935nm pulse with 860MHz sidebands mentioned in the text.}
\label{fig:Popuations}
\end{figure}

We demonstrate this exponential suppression of leaked population using a $^{171}$Yb$^+$ ion in an RF Paul trap employing standard Doppler cooling, state initialization and read-out schemes\cite{Olmschenk07}. The measurement is made by preparing one of the leakage states, applying $n$ cycles of the leakage repump protocol, and then reading out the populations of the four different states in $^2S_{1/2}$, (Fig.~\ref{fig:Popuations}). State initialization is performed by optically pumping to $\ket{0}$, followed by a 20$\mu s$ microwave $\pi$-pulse at $\nu_0\pm\nu_Z$, where $\nu_0\approx12.643$GHz is the qubit splitting in a magnetic field of 5.6G and $\nu_Z=$7.8MHz is the Zeeman splitting of the $m_f=\pm1$ states. The quadrupole transition is driven with $8$mW of 435nm light, focused to a beam diameter of 50$\mu$m, resulting in a ~$1\mu$s $\pi$-time. The $\ket{L_{\mp}}\rightarrow{}^2D_{3/2}\ket{F=1,m_f=\pm1}$ transfer pulses are done sequentially for convenience, but could in principle be done simultaneously. Before measurement, we apply an additional 935nm pulse with 860MHz sidebands to clean out $^2D_{3/2}\ket{F=2}$ population stemming from the $1.8\%$ branching ratio of $^3[3/2]_{1/2}$ and off-resonant coupling during the quadrupole transition. However, this population accrual should be small as it is already mitigated during the sequence due to power broadening of the $^2D_{3/2}\leftrightarrow{}^3[3/2]_{1/2}$ transition. Standard state-dependent fluorescence read-out of the $^{171}$Yb$^+$ qubit mixes the three states in the $F=1$ manifold\cite{Olmschenk07}, providing an estimate of the $F=0$ population and a sum of the populations in $F=1$. We measure the population in a single $\ket{F=1,m_f}$ state by applying a final microwave pulse to swap $\ket{0}$ with the population we want to measure. As shown in Fig.~\ref{fig:Popuations}, we observe an exponential decay in the leakage poplulation with a slight deviation from the ideal decay of $1/3^n$ due to imperfect transfer pulses subject to Debye-Waller effects\cite{wineland1998}, laser and B-field fluctuations, and imperfect 935nm polarization that drives population to $^3[3/2]_{1/2}\ket{F=1,m_f=0}$ which decays into the qubit subspace with probability 1/3.

Deviations from the ideal pumping scheme are quantified by fitting the observed data with a simple model. Ignoring coherent effects, each repumping cycle will change the population distribution according to a linear operator $R$, such that the population after the $n^{th}$ cycle is $P^{(n)}=RP^{(n-1)}$. We order the populations so that $\{P^{(n)}_1,P^{(n)}_2,P^{(n)}_3,P^{(n)}_4\}=\{P^{(n)}_0,P^{(n)}_{L_-},P^{(n)}_1,P^{(n)}_{L_+}\}$ and $R_{ij}$ is the probability of state $j$ getting pumped to state $i$ in a single cycle. Observing that the qubit states are approximately steady states, we set $R_{i,1}=\delta_{i,1}$ and $R_{i,3}=\delta_{i,3}$. We also assume a symmetry in the leakage states so that $R_{2,2}=R_{4,4}$, $R_{2,1}=R_{4,1}$ and $R_{2,3}=R_{4,3}$. These assumptions and the normalization $\sum_jR_{i,j}=1$ reduce the model to three free parameters, $R_{2,1}$, $R_{2,2}$ and $R_{2,3}$ allowing for a straight-forward calculation of the populations after $n$ pulses as $P^{(n)}=AR^nP^{(0)}+B$. We've included two additional fitting constants, $A$ and $B$, to account for state-intialization and measurement errors with their ideal values being $1$ and $0$ respectively. As shown in Fig.~\ref{fig:Popuations}, the data is in relatively good agreement with the fit parameters being $\{R_{2,1},R_{2,2},R_{2,3},A,B\}=\{0.323(9),0.27(2),0.225(5),0.952(6),0.026(1)\}$. Ideally $R_{2,1}=R_{2,2}=R_{2,3}=1/3$ and our measurement showing an asymmetry in the pumping rates into the two qubit states and the transient transfer of population from $\ket{L_-}$ to $\ket{L_+}$ implies the presence of polarization impurities in the 935nm beam.

An important requirement of the leakage repump is that it leave qubit population unperturbed. The quadrupole transfer pulses can induced new leakage errors by one of two means: 1) off-resonant coupling from $\ket{1}$ to $^2D_{3/2}\ket{F=2,m_f=\pm2}$ or 2) polarization or k-vector misalignment driving $\Delta m_f=0,\pm1$ transitions. The narrow linewidth of the quadrupole transition allows these errors to be controlled via pulse shaping or longer $\pi$-times. For a square pulse, the first mechanism results in an induced error per cycle of $(\Omega_{\epsilon}/\delta_{hf})^2$ where $\delta_{hf}/2\pi=860$MHz is the hyperfine splitting of $^2D_{3/2}$ and $\Omega_{\epsilon}$ is the on-resonance Rabi frequency for $\ket{1}\leftrightarrow{}^2D_{3/2}\ket{F=2,m_f=\pm2}$. We relate this to the $\pi$-time of the transfer pulse $\tau_{\pi}=\pi/\Omega_0$ by noting that $|\Omega_0/\Omega_{\epsilon}|=3/2\sqrt{2}$, so that the induced error per cycle is $\approx8\pi^2/(3\delta_{hf}\tau_{\pi})^2<10^{-6}$ when $\tau_{\pi}=1\mu$s. Assuming the polarization and k-vector can be aligned to approximately $1^{\circ}$, $\Delta m_f=\pm1$ transitions will induce a similar in magnitude error, and the $\Delta m_f=0$ transitions have a negligible contribution. We mitigate this error even further by rounding the pulse edges with $700$ns turn-on and turn-off times. However, even with pulse shaping, the unwanted transition will always be transiently populated and the ultimate limit for this error is $\approx\gamma\tau_{\pi} (\Omega_{\epsilon}/\delta_{hf})^2$, where $\gamma=(52.7ms)^{-1}$ is the scattering rate of $^2D_{3/2}$. For $\tau_{\pi}=1\mu$s, we estimate a fundamental error of ~$10^{-11}$ per cycle.

The leakage repump induces a differential AC Stark shift on the qubit $\delta\omega\approx\Omega_{\epsilon}^2/2\delta_{hf}$, resulting in a phase shift of $\delta\omega\tau_{\pi}=(4\pi^2)/(9\delta_{hf}\tau_{\pi})\approx 1$mrad for $\tau_{\pi}=1\mu$s. This shift could result in a $\sim10^{-6}$ induced error, but we do not consider this a fundamental error since it can easily be corrected by a $\hat{Z}$ rotation.

\begin{figure}
  \centering
    \includegraphics[width=0.45\textwidth]{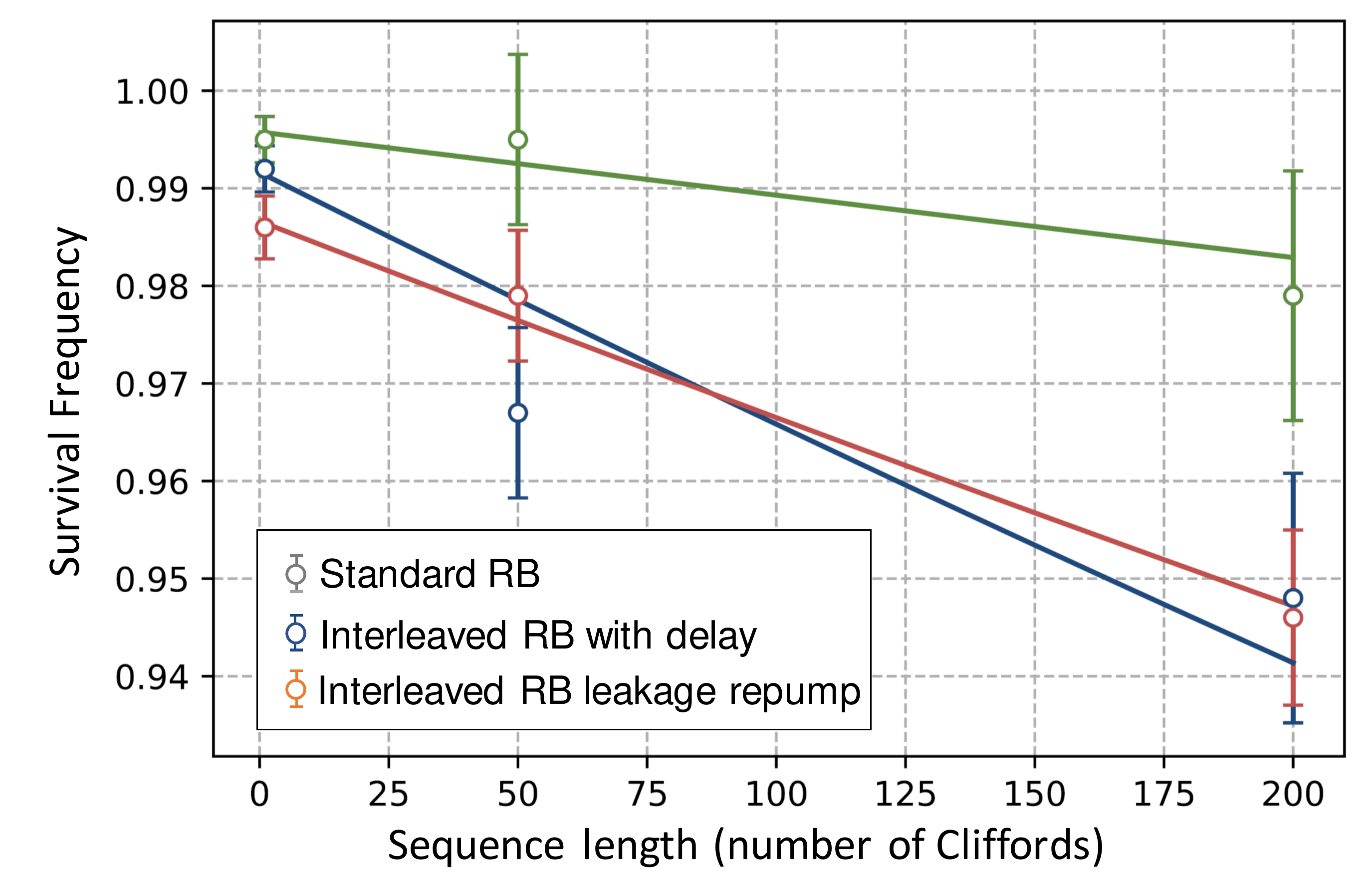}
\caption{Benchmarking the identity operation with and without the leakage repump protocol being implemented. The survival frequency of randomized benchmarking quantifies the probability of a random sequence generating a target state and decays exponentially as errors accumulate. This decay is fitted to a function that is related to the average fidelity, which is used to bound the induced error from the leakage repump protocol. We observe no significant increase in the error rate when the leakage repump sequence is implemented during the benchmarked identity operation, meaning that the measurement is limited by memory error. We, therefore, upper bound the induced error at $\leq 2.0(8)\times10^{-5}$ per cycle.}
\label{fig:RB}
\end{figure}

 Errors induced by the leakage repump protocol on the qubit subspace can be quantified via interleaved randomized benchmarking (IRB) experiments\cite{Magesan12}. IRB is a protocol designed to quantify an error rate for a particular gate of interest. The basic idea of IRB is that a particular gate's error can be quantified by repeatedly inserting it into randomized gate sequences and subtracting off the error ascribed to the truly random component. Presuming that the leakage repump protocol would be implemented during idle times, (during the identity gate), we benchmark the identity gate both with and without the leakage repump protocol being simultaneously performed.  Our IRB measurement consists of two experiments: first, standard randomized benchmarking, which acts as a reference, and second, randomized benchmarking plus an $n=10$ leakage repump sequence applied after each gate. We also ran IRB with a similar delay time interleaved in the second experiment to measure the memory errors (similar to Ref.~\cite{Sepiol19}).  For each IRB run, we measure the survival probability of ten random sequences for three different sequence lengths. The results of each experiment are plotted in Fig.~\ref{fig:RB} along with fits to the standard decay equation. The decay rate of each experiment is then used to estimate the interleaved gate's average error, $\epsilon_g = \tfrac{1}{2}(1-p_2/p_1)$, where $p_{1/2}$ are the decay rates from the first and second experiments for each IRB run. From IRB, we estimated the average $n=10$ leakage repump error is $\epsilon_{\textrm{repump}}^{n=10} =2.0(8)\times10^{-4}$ and the average memory error is $\epsilon_{\textrm{memory}} =1.5(6)\times10^{-4}$ with the uncertainties estimated from confidence intervals obtained by semi-parametric bootstrap resampling~\cite{Meier06}. These measurements imply that the effects of leakage repump are dominated by the idling memory errors of the system and we are, therefore, only able to establish an upper bound for the error per cycle of $\epsilon_{\textrm{repump}}^{n=1}\leq2.0(8)\times10^{-5}$.

\begin{figure}
  \centering
    \includegraphics[width=0.5\textwidth]{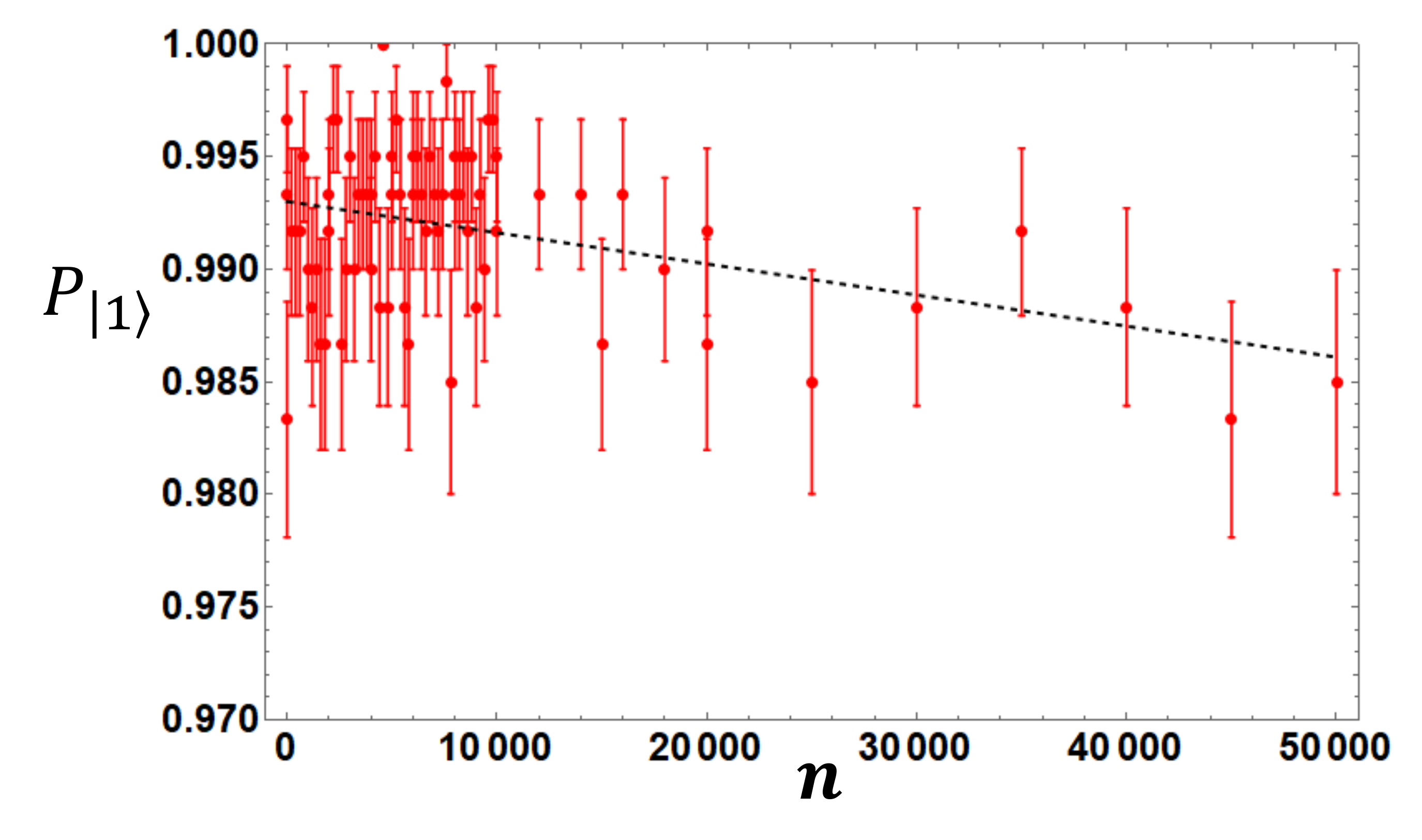}
\caption{Population decay within the qubit subspace during the leakage repump protocol. The measurement is made by preparing $\ket{1}$, applying the leakage repump cycle $n$ times and then measuring the $\ket{1}$ population. Fitting a decaying exponential to the data results in a decay per cycle of $1.4(3)\times10^{-7}$, which is depicted by the black dashed line.}
\label{fig:T1}
\end{figure}

Since our IRB measurements are limited by intrinsic memory errors of the system, we also measured the population decay out of $\ket{1}$ as a function of the number of repump cycles. As shown in Fig.~\ref{fig:T1}, fitting the data to a decaying exponential results in a decay constant of $1.4(3)\times10^{-7}$ per cycle.

A rough analysis of the protocol can be made by assuming a qubit initialization procedure with an error $\epsilon_0$ dominated by a leakage inducing spontaneous emission process. Just how much leakage suppression is needed depends on the application, but as a case study, we examine implementing quantum error correction via the distance-3 surface-code. An experimental demonstration of the break-even point, where the logical error rate is equal to the physical error rate, would require $\epsilon_0=\mathcal{O}(10^{-3})$\cite{Tomita14}. A distance-3 code can correct a single error, but will fail if two errors occur. If a leakage event occurs during a two-qubit gate in the syndrome extraction, one qubit will leak and the other will partially depolarize. Avoiding a subsequent error when the leaked qubit interacts with another data qubit requires the leakage population to be suppressed back below the threshold error level. We therefore require a repump cycle that provides a $10^3$ level of suppression after each gate. We define the probabilities of errors induced by each repump cycle as $\epsilon_l$ and $\epsilon_q$ where the subscripts respectively denote leakage-inducing and non-leakage-inducing errors. Assuming the ideal pumping rate of $1/3^n$, the leakage population after $n$ cycles is $\epsilon_0(1/3)^n+3\epsilon_l(1-1/3^{n})/2$ and the total error is $\epsilon_0+n(\epsilon_q+\epsilon_l)$. Using our induced error upper bound of $2.0(8)\times10^{-5}$ and our population decay measurement that implies $\epsilon_l<<\epsilon_q$, we find that an $n=7$ sequence would provide the necessary leakage suppression, only increase the total error by $1.4\times10^{-4}$, and take approximately $70\mu$s for our experimental parameters. However, given more laser power, further optimizations such as applying the two quadrupole drives simultaneously and more advanced pulse-shaping would enable shorter times ultimately limited by the $38$ns decay time of $^3[3/2]_{1/2}$. We estimate that the required leakage suppression could be done in $<10\mu$s, which is small compared to typical gate and cooling times.

Our protocol can be generalized for use in atoms with more complicated hyperfine structures. The $I=1/2$ nucleus in $^{171}$Yb$^+$ simplifies the scheme by ensuring that there are no $\Delta m_f=\pm2$ transitions for the qubit states in $^2D_{3/2}\ket{F=1}$, and $I>1/2$ systems will need to rely on spectroscopic resolution to suppress excitations out of the qubit space. On the other hand, these systems admit clock qubits at higher fields, resulting in larger energy splittings and looser spectroscopic resolution requirements. While a procedure suited for $I>1/2$ systems would still exponentially suppress leakage population, it is not likely to achieve the same efficiency due to the larger number of decay channels. A generalized protocol would also be more complex since it would require more laser frequency tones.

In conclusion, our protocol supresses leakage population by orders of magnitude with a neglible induced error in the qubit space. While we obviate the need for qubit or circuitry overheads to deal with leakage, we do so at the expense of introducing another laser beam for the quadrupole transition. However, this additional hardware complexity to suppress leakage at the physical level not only preserves an error correcting code's performance, but could also be faster than algorithmic methods. We also note that actively removing leakage to a certain level is superior to a gate designed with an equivalent inherent leakage level, since in the latter case leakage errors will continue to accumulate and spoil longer algorithms. This work removes a significant obstacle to the error mitigation techniques that will be crucial to both near and long term development of trapped-ion quantum computers.

\begin{acknowledgments}
The authors would like to thank Michael Foss-Feig for helpful comments on an earlier version of this manuscript.
\end{acknowledgments}

\end{document}